\documentclass[useAMS,usenatbib]{mn2e}
\usepackage{amssymb}
\usepackage{graphicx}

\title[Photopolarimetric study of CB56, CB60, and CB69]{Study of background star polarization and polarization efficiency of three selected Bok globules CB56, CB60 and CB69}
\author[A. Chakraborty H. S. Das and D. Paul]
{A.Chakraborty$^{1}$\thanks{E-mail: carindam1@gmail.com (AC)}
,H. S. Das$^{1}$\thanks{E-mail: hsdas@iucaa.ernet.in (HSD), corresponding author}
,D. Paul$^{2}$\thanks{E-mail: dipankarrkpaul99@gmail.com (DP)}\\
\\
$^{1}$Department of Physics, Assam University, Silchar 788011, India\\
$^{2}$Department of Physics, Ramkrishna Nagar College, Ramkrishna Nagar 788166, India}
\begin{document}

\date{Accepted 2014 April 15. Received 2014 April 15; in original form 2013 October 17}

\pagerange{\pageref{firstpage}--\pageref{lastpage}} \pubyear{2014}

\maketitle

\label{firstpage}

\begin{abstract}
We present the polarization maps of three selected Bok globules CB56, CB60 and CB69 constructed using a V-band data from a CCD imaging polarimeter. The aim of this work is to measure the optical polarization
($p_v$) of background field stars in order to determine the polarization efficiency, $p_v/A_v$. We find that the local magnetic field of the cloud CB56 is almost aligned with the galactic field, but not in CB60 and CB69.  A trend of decreasing polarization efficiency with increasing extinction ($A_v$) is observed: it can be well represented by a power law, $p_v/A_v \propto A_v^{-\alpha}$, where $\alpha = -0.56 \pm 0.36$, $-0.59 \pm 0.51$ and $-0.52 \pm 0.49$ for CB56, CB60 and CB69 respectively. This indicates that the linear polarization of the starlight
due to aligned dust grains in these clouds is produced more efficiently in low extinction regions, compared with high obscured lines of sight.

\end{abstract}


\section{Introduction}

Bok globules are the most suitable sites for studying the evolution and formation of isolated low mass stars    \citep{b3}. Aligned dust grains in the interstellar medium (ISM) and the Bok globule polarize the radiation through dichroic extinction. The grains are aligned with their short axis preferentially aligned with the local magnetic field, resulting in a polarization parallel to the field, as projected onto the plane of the sky (Lazarian et al. 1997).

 Optical polarimetric observations of dark cloud can give information about the magnetic field orientation in the low density edge regions of clouds, whereas the polarimetric observation in the infrared and submillimetre range can map the field orientation in the higher density central regions of the clouds. Thus the resulting polarization map outlines the geometry of the magnetic field lines projected onto the plane of the sky, which in turn helps to study the  relationship between structure kinematics, and embedded magnetic field during various stages of star formation (\citet{bvss},\citet{ bv},\citet{ bvls},Vrba, Coyne \& Tapia (1981), \citet{bgj}, \citet{b22}, \citet{bbj}, \citet{b11}, \citet{b13}, \citet{bafg}, \citet{bwskn}, \citet{bfag}).

\cite{jtj} and \cite{jkd} modeled the observed variation in polarization with extinction along different lines of sight, in terms of a magnetic field with distinct random and uniform components. \cite{jkd} argued that most  of the observed fall-off in $p/A_V$ is due to the turbulence of the magnetic field in the medium.  \citet{jtj} predicted that $p/A$ $\propto$ A$ ^{-0.5}$, which is closed to the observed dependence. It has been investigated later that Jones model may not be applicable to individual clouds. \citet{b10} studied the dust grain alignment in the Taurus dark cloud (TDC) and showed a strong systematic trend in polarization efficiency with extinction which can be expressed by a power law $p/A$ $\propto$ A$ ^{-0.56} $, in the magnitude region (0 $ \leq A_{v} \leq $ 25). They explained this phenomenon of decrease in polarization with increase in extinction by considering a number of factors (poor grain alignment, grain growth, and or changes in grain shape or composition).  \citet{bwth} also did a detailed study of interstellar polarization efficiency toward molecular clouds to understand the grain alignment mechanism in different regions of interstellar medium. They studied the dependence of polarization efficiency on  visual extinction by probing background field stars, and found that this dependency can be well expressed by a power law $p_{\lambda}/\tau_{\lambda}$ $\propto$ $A_{v} ^{-0.52} $ ($ \tau_{\lambda} $ is the optical depth). \citet{bwth} also showed that, among all the available grain alignment models, the radiative torque mechanism could more efficiently explain the observed polarization in the quiescent regions of the cloud.

  In this work, we have probed CB56, CB60 and CB69 at the V-band, with a scientific goal to measure the optical polarization of the stars background to the globules and to map the magnetic fields within these globules. The extinction of field stars in each cloud are estimated using a standard technique suggested by \cite{b2} and \cite{b141}. The extinction of each field stars in the cloud and background star polarization will help to understand the dynamical behavior of the cloud.  Some works on cloud CB56 and CB60 were reported by a few investigators. Both near infrared  imaging and millimetre dust emission of CB60 were studied by \citet{byc} and \citet{blh} respectively and found that it is about 1500 parsec(pc) away from the sun.  \citet{blh}  also studied the millimetre dust emission of CB69 and reported a distance of 500pc. This is only marginally smaller than the distance of 600 pc derived by \citet{b4}. Further, \citet{b12} mapped the magnetic fields within CB56 at R-band.

This article is arranged as follows. In Section 2 we discuss the observational procedures and details of the data reduction techniques are discussed in Section 3. This is followed by results and discussion in Section 4; we end with appropriate conclusions in Section 5.

\section[]{OBSERVATION }
\subsection{Polarimetric data}
Polarimetric observation of three Bok globules CB56, CB60 and CB69, have been made at the V-band with exposure time of 900 seconds, using 2-m Telescope of Girawali Observatory at Inter University Center for Astronomy and Astrophysics (IUCAA), Pune (IGO, Latitude: 19$^\circ $5$'$ N, Longitude: + 73$^\circ $40$'$ E, Altitude = 1000 m), India on 4th and 5th March, 2011. These globules were selected from the catalogue of \citet{b4}, depending on their visibility at the time of observation. The observational details of the field containing three Bok globules, are listed in Table 1 (columns 1, 2 and 3 give serial number, globule identification and position angle (in degree), respectively; columns 4 and 5 give 2000.0 epoch right ascension (RA) and declination (Dec.), respectively; columns 6 and 7 give the Galactic coordinates, respectively; column 8 gives the dimension of the globules and columns 9 and 10 gives the date of observation and UT, respectively).

\begin{table*}
 \begin{center}
   \begin{minipage}{140mm}
  \caption{Observational log of Bok globules.}
  \begin{tabular}{|c|c|c|c|c|c|c|c||c|c|c|c|c|c|}

\hline
Serial  & Object  & Position Angle & RA(J2000)  & Dec.(J2000) &  \textit{l}
        & \textit{b} & Dimension & Date & UT  \\
number  & ID       & (in degree) & (h m s)  &  $ (^\circ $ $^{\prime}$  $ ^{\prime\prime} $)
        &   $(^\circ)$ &  $(^\circ)$ & ~ & ~ & (h m)\\

\hline

 1 & CB56 & 170 & 07 14 36 & -25 08 54.20& 237.93 & -6.46 &  4.5$'$ x 2.2$'$ & 04/03/2011 & 17 23  \\
 2 & CB60 &  40 & 08 04 36 & -31 30 47.00& 248.89 & -0.01 & 10.1$'$ x 6.7$'$ & 05/03/2011 & 17 38  \\
 3 & CB69 & 120 & 17 02 42 & -33 17 00.00& 351.23 &  5.14 & 10.1$'$ x 4.5$'$ & 05/03/2011 & 22 32  \\
\hline

\end{tabular}
\label{tab1}
\end{minipage}
\end{center}
\end{table*}

\subsection{Photometric data}
Photometric (BVR) data of the field stars of CB56, CB60 and CB69 have been obtained from the VizieR data base of astronomical catalogues, namely the Fourth US Naval Observatory CCD Astrograph Catalog (UCAC4: \citet{b18}), the Naval Observatory
Merged Astrometric Dataset (NOMAD: \citet{b17}), and the Deep Near Infrared Survey of the Southern Sky (DENIS) data
base (DENIS: \citet{b19}).  The photometric magnitudes of
each field star of known coordinates (RA (J2000) and Dec. (J2000))
have been extracted via the VizieR service of Centre de Donn´ees
astronomiques de Strasbourg (CDS), using 2-arcmin search radii as
query from these catalogues.
\subsection{Instrumentation}

The IGO 2-m telescope has a Cassegrain focus with a focal ratio of f/10. The IUCAA Faint Object Spectrograph and Camera (IFOSC) is the main instrument available on the telescope's direct Cassegrain port, for the observation. The imaging have been done using an EEV 2K $\times$ 2K pixel$^2$ thinned, back-illuminated CCD which provides an effective field of view of 10.5 arcmin radius and each pixel corresponds to 0.3 arcsec on the sky. IFOSC$'$s capabilities are enhanced with an imaging polarimetric mode (IMPOL) with a reduced circular field of view of about 2 arcmin radius. It measures linear polarization in the wavelength band: 350$-$850nm. The design of the IMPOL polarimeter is standard with a stepped half wave plate (HWP) followed by a Wollaston prism; a focal mask is used to prevent the ordinary and extra ordinary images overlapping. The instrument have an built-in acquisition and guidance unit. The detail description of the instrument is given in \citet{b16}.

\section{DATA REDUCTION AND CALIBRATION}

\subsection{Polarimetry}
A ratio R($\alpha$) is defined as;\\
\begin{equation}
R(\alpha)=\frac{I_{o}/I_{e}-1}{I_{o}/I_{e}+1}= \textit{p}\cos(2\theta-4\alpha) \\
\end{equation}

Here $I_{e}$ and $I_{o}$ are extraordinary and ordinary image, $ \textit{p} $ is the fraction of the total light in the linearly polarized condition, $\theta$ and $\alpha$ are the position angles of the polarization vector and the half-wave plate fast-axis respectively, with reference to the axis of the Wollaston prism. Since the angle $\theta $ is conventionally measured eastward from north (0$ ^{\circ} $ towards  celestial north pole and increasing counter-clockwise), the axis of the Wollaston prism is kept aligned to it. The ratio R($\alpha$) is related  to the normalised Stokes parameters as follows:\\

\begin{equation}\label{eq2}
R(0^\circ)=-R(45^\circ)=q ; ~~~ R(22.5^\circ)=-R(67.5^\circ)=u\\
\end{equation}

The linear polarization components, (q=Q/I) and (u=U/I), of a target can be obtained from measurements taken at four position angles of the HWP of 0$ ^{\circ} $, 45$^{\circ}$, 22.5$^{\circ}$, and 67.5$^{\circ}$, \\
\\(I=I$ _{e}(\alpha)+I _{o}(\alpha) $).\\

The linear polarization $ \textit{p} $ and position angle $\theta$ of the polarization vector is given by;

\begin{equation}\label{eq3}
 \textit{p} =\sqrt{q^{2}+u^{2}}~~~ and~~~ \theta=\tan^{-1}(q/u)
\end{equation}

Both $\textit{p}$ and $ \theta $ can be determined by using only two Stoke$'$s vectors q and u. But, the polarimetric measurements of a target  in additional two rotation i.e. at ($\alpha$ = 45$^{\circ}$ and 67.5$^{\circ}$) are taken to counter the non-responsivity of the system \citep{b16}.

\subsection{Image processing}
The data reduction procedures and polarization  measurement are similar to those described in \citet{bdmwbdc}. Standard IRAF routines have been used for bias-subtraction, and flat-field corrections of the CCD images of CB56, CB60 and CB69, and for the polarimetric standard stars. From the cleaned science images at four position angles of HWP, the linear polarization of an object can be obtained as follows.

 Aperture photometry in each frame have been carried out to estimate the magnitude of both the ordinary and extraordinary images of an object, using a task PHOT in APPHOT package within IRAF. The aperture should be taken such a way that the noise from the background is minimized and the signal from an object is maximized. The radius of the aperture has been chosen within two to three times of FWHM, so that the polarization value of a star has a minimum error in polarization, where FWHM is $\sim$ 2 arc sec.

The linear polarization of an object has been calculated using equation (3).  The errors in the polarization measurements are primarily dominated by photon noise. For low values of polarization, the number of photo-electrons corresponding to ordinary and extraordinary images are approximately equal. The estimate of error in the measurement of $ \textit{p} $ and  $\theta$ based on photon noise is given by (\citet{b16}):\\
\begin{equation}
\sigma_{\textit{p}}=\frac{\sqrt{N+N_{b}}}{N}~~~and~~~\sigma_{\theta}= 0.5\times\frac{\sigma_{\textit{p}}}{\textit{p}}~~rad
\end{equation}
where N and N$_{b}$ are the flux counts from star and the background respectively.

\subsection{Polarimetric calibration}

\subsubsection{Instrumental polarization and position angle offset}
The instrumental polarization are corrected using unpolarized standard stars, having very small polarization, typically $\emph{p} $ $\textless$ 0.05  $\%$. The offset in the position angle between the celestial and instrumental coordinates are corrected by using strongly polarized standard stars.  We have observed both unpolarized (HD-65583) and polarized (HD-43384, HD-251204 and HD-147084) standard stars for calibration purpose. The standard stars for
null polarization and zero-point of the polarization position angle (PA) were taken from Serkowski et al. (1975) and HPOL\footnote {http://www.sal.wisc.edu/HPOL/tgts/HD251204.html}, respectively. Table \ref{tab2} and \ref{tab3}  show the typical instrumental polarization and the offset in polarization position angle, which is (${\textit{p}_{\circ}})_{av}= -0.08\%$ and  $(\theta_{\circ})_{av} = -0.68^{\circ}$, respectively. These two values are used to calibrate our results.

\begin{table*}
 \begin{center}
  \begin{minipage}{140mm}
  \caption{Observational result of unpolarized standard stars. Here the values listed under the column $p$, $\theta$ are taken from Serkowski et al. (1975), $p_{obs}$ is the observed results. Offset in polarization is given by $\textit{p}_{\circ}= (\textit{p} - \textit{p}_{obs}$).}
  \begin{tabular}{|c|c|c|c|c|c|c|c||c|c|c|c|c|}

\hline
Serial  & Star  & Date & Filter &$\textit{p}$ &  $ \theta $ & $\textit{p}_{obs} $ & ~ &
   $ \textit{p}_{\circ}$   \\
 number  & ~   & ~ & ~&($\%$) & ($^{{\circ}}$) &($\%$) & ~& ($\%$)  \\
\hline
1 & HD-65583 & 04/03/2011 & V & 0.01  & 144.70 & 0.09$ \pm $0.07& ~& $-0.08$   \\

 \hline
2 & HD-65583 & 05/03/2011&  V & 0.01  & 144.70 & 0.09$ \pm $0.06 & ~& $-0.08$  \\

\hline
\end{tabular}
\label{tab2}
\end{minipage}
\end{center}
\end{table*}
\begin{table*}
 \begin{center}

 \begin{minipage}{140mm}
  \caption{Observational result of polarized standard stars. Here the values listed under the column $p$ and $\theta$ are taken from Serkowski et al. (1975) and  HPOL, while the values listed under the column $p_{obs}$ and $\theta_{obs}$ are our observed results. Offset in PA is given by $ \theta_{\circ}$= ($\theta - \theta_{obs}$).}
  \begin{tabular}{|c|c|c|c|c|c|c|c||c|c|c|c|c|}
\hline
 Serial  & Star  & Date & Filter &$\textit{p}$ &  $ \theta $ & $\textit{p}_{obs} $ & $\theta_{obs} $ & $\theta_{\circ} $\\
 number  & ~  & ~ & ~&($\%$) & ($^{{\circ}}$) &($\%$) & ($^{{\circ}}$) & ($^{{\circ}}$) & \\
 \hline
1 & HD-43384 & 04/03/2011 & B & 2.83$ \pm $0.05 & 169.3$ \pm $0.7& 2.77$ \pm $0.07 & 171.5$ \pm $0.7  & --2.2 \\
   &          &            & V & 2.94$ \pm $0.04 & 169.8$ \pm $0.7& 2.90$ \pm $0.07 & 171.9$ \pm $0.7 & --2.1  \\
   &          &            & R & 2.86$ \pm $0.03 & 170.7$ \pm $0.7& 2.88$ \pm $0.08 & 169.5$ \pm $0.8 &  1.2 \\
 \hline
 2 & HD-251204& 04/03/2011& B & 4.46 & 155.8 & 4.83$ \pm $0.09& 155.9$ \pm $0.5 & --0.1\\
   &          &           & V & 4.72 & 153.3 & 4.93$ \pm $0.06& 154.2$ \pm $0.3 & --0.9\\
   &   		  &	          & R & 4.80 & 154.8 & 4.88$ \pm $0.10& 152.9$ \pm $0.6 &  1.9\\
 \hline
 3& HD-147084 & 05/03/2011 & B & 3.50 & 32.0& 3.43$ \pm $0.10 & 33.6$ \pm $0.8  & --1.6 \\
  &		   &               & V & 4.18 & 32.0& 4.08$ \pm $0.09 & 33.7$ \pm $0.6  & --1.7 \\
  &           &            & R & 4.44 & 32.2& 4.32$ \pm $0.10 & 32.8$ \pm $0.6  & --0.6 \\
\hline
\end{tabular}
\label{tab3}
\end{minipage}
\end{center}
\end{table*}

\section{RESULTS AND DISCUSSION}
\subsection{Linear polarization}
The details of the observed linear polarization of the field stars of three Bok globules CB56, CB60, and CB69 are given in Table \ref{tab4}, \ref{tab5}, and \ref{tab6}, where column one gives the serial number of the stars; column two and three give RA (J2000) and Dec (J2000) coordinates; column four and five give the observed polarization and position angle, column 6, 7, 8 give the (BVR) magnitude and column 10 comments whether the star belongs to main sequence.

\begin{figure*}
\vspace{27pt}
\begin{center}
 \includegraphics[width=25pc, height=20pc]{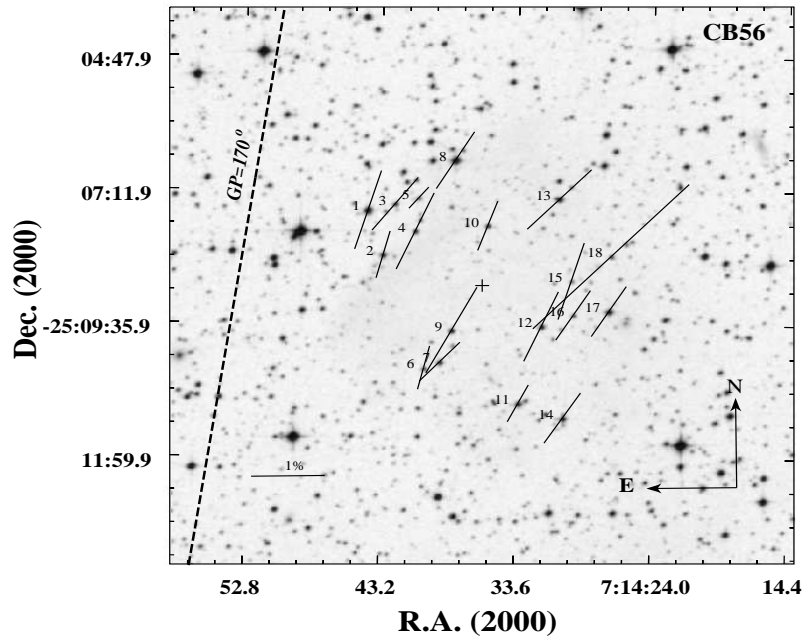}
\caption{The stellar polarization vectors and position angles are superimposed on a 10$'$ x 10$'$ R-band DSS image of the field containing CB56.  A vector with a polarization of 1$\%$ is drawn for reference, the length of all the polarization vector is proportional with it. The dashed line represents the projection of Galactic plane at b = $-6.46^{\circ}$ which corresponds to a position angle of 170$^{\circ}$. The `+' symbol represents the globule center RA = 07h 14m 36s and Dec = $-$ 25d 08m 54.20s.}
\label{Fig1}
\end{center}
\end{figure*}

\begin{figure*}
\begin{center}
\vspace{25pt}
 \includegraphics[width=25pc, height=20pc]{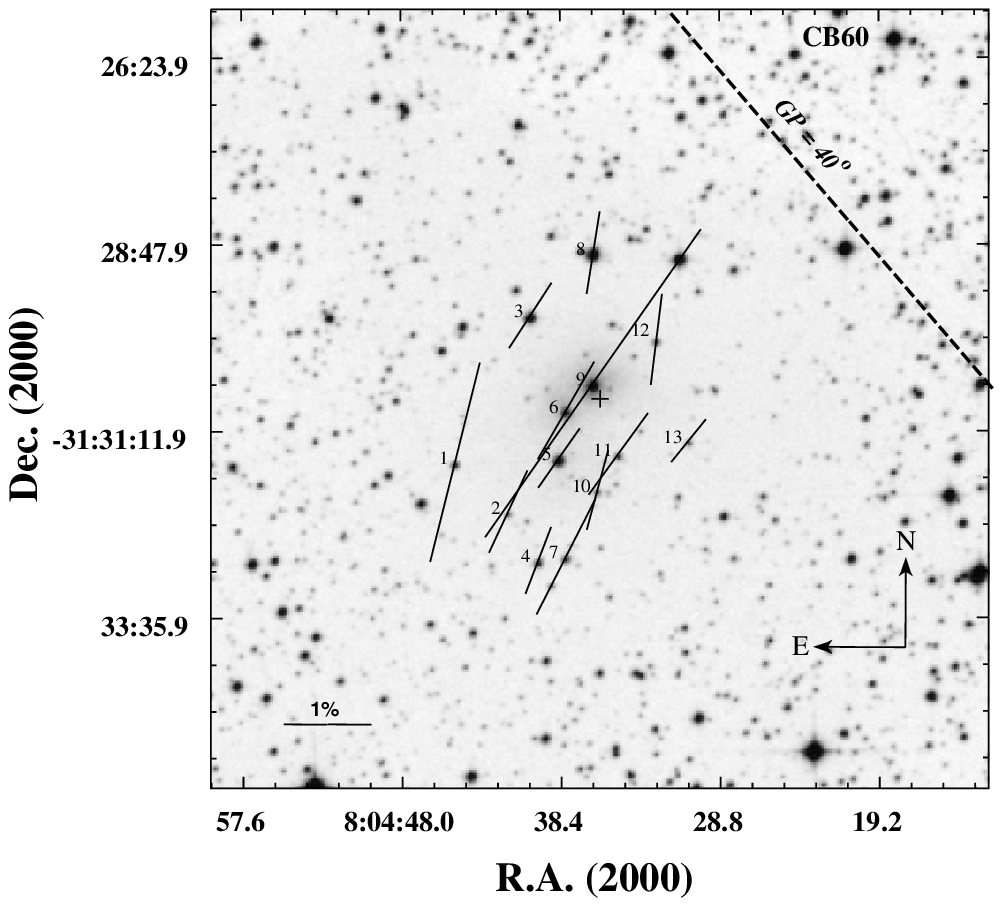}
 \caption{The stellar polarization vectors and position angles are superimposed on a 10$'$ x 10$'$ R-band DSS image of the field containing CB60.  A vector with a polarization of 1$\%$ is drawn for reference, the length of all the polarization vector is proportional with it. The dashed line represents the projection of Galactic plane at $b = -0.01^{\circ}$ which corresponds to a position angle of 40$^{\circ}$. The `+' symbol represents the globule center RA = 08h 04m 36s and Dec = $-$31d 30m 47.00s.}
\label{Fig2}
\end{center}
\end{figure*}

\begin{figure*}
\begin{center}
\vspace{25pt}
\includegraphics[width=25pc, height=20pc]{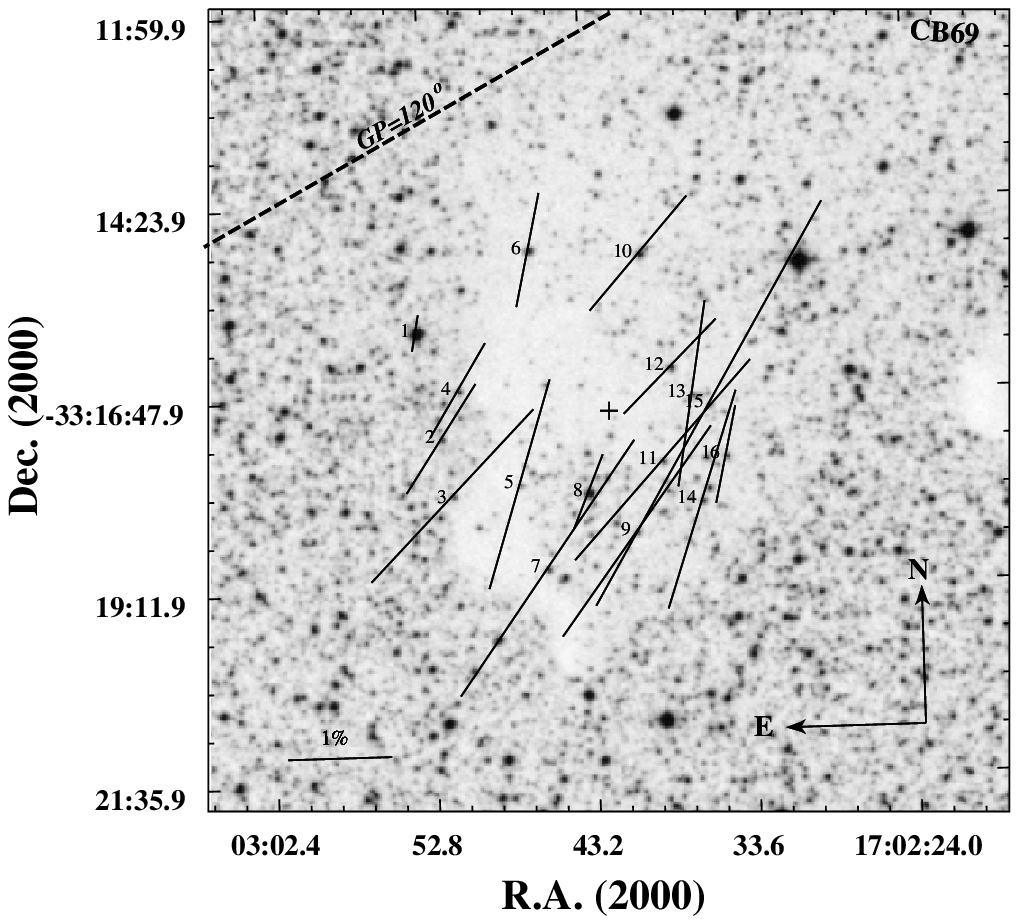}
\caption{The stellar polarization vectors and position angles are superimposed on a 10$'$ x 10$'$ R-band DSS image of the field containing CB69.  A vector with a polarization of 1$\%$ is drawn for reference, the length of all the polarization vector is proportional with it. The dashed line represents the projection of Galactic plane at $b = 5.14^{\circ}$ which corresponds to a position angle of 120$^{\circ}$. The `+' symbol represents the globule center RA = 17h 02m 42s and Dec = $-$33d 17m 00.00s.}
\label{Fig3}
\end{center}
\end{figure*}

\begin{table*}
\begin{center}
  \caption{Estimated values of linear polarization in V-band ($\textit{p}_{v}$), polarization angle ($\theta $) and BVR magnitude of various field stars in CB56 (NA: Not Available).}
  \begin{tabular}{|c|c|c|c|c|c||c|c|c|c|}
  \hline
Serial   & RA(J2000)  & Dec.(J2000) &  $\textit{p}_{v}$ & ($\theta $)& B& V  &  R & Main sequence star \\
number   & (h m s)    & $ (^\circ $ $^{\prime}$  $ ^{\prime\prime} $) & $(\%)$ &  $(^\circ)$&(mag)&(mag)&(mag)&(yes/no) \\
\hline%
  1&   07  14    44.14 &-25   07   33.71 &    1.25
$\pm$     0.46 & 163 $\pm$  11
&   14.98$\pm$    0.06&   13.70$\pm$    0.01&   13.27$\pm$    0.01
&  yes \\
  2&   07  14    43.04 &-25   08   21.65 &    0.74
$\pm$     0.27 & 165 $\pm$  10
&   16.49$\pm$    0.07&   15.29$\pm$    0.01&   14.76$\pm$    0.01
&   yes \\
  3&   07  14    42.23 &-25   07   26.93 &    1.02
$\pm$     0.85 & 142 $\pm$  24
&   16.80$\pm$    0.01&   15.64$\pm$    0.01&   15.38$\pm$    0.01
&  yes \\
  4&   07  14    40.79 &-25   07   56.07 &    1.27
$\pm$     0.31 & 156 $\pm$   08
&   16.04$\pm$    0.01&   15.55$\pm$    0.01&   15.56$\pm$    0.01
&  yes \\
  5&   07  14    40.56 &-25   07   20.17 &    0.42
$\pm$     1.01 & 140 $\pm$  69
&   16.96$\pm$    0.09&   16.33$\pm$    0.01&   15.80$\pm$    0.02
&   no\\
  6&   07  14    40.08 &-25  10   23.33 &    0.68
$\pm$     0.75 & 166 $\pm$  32
&   17.15$\pm$    0.02&   16.59$\pm$    0.02&   15.63$\pm$    0.02
&  no\\
  7&   07  14    38.95 &-25  10   16.70 &    0.81
$\pm$     0.20 & 137 $\pm$   07
&   15.97$\pm$    0.12&   15.17$\pm$    0.07&   14.97$\pm$    0.04
&   yes\\
  8&   07  14    38.00 &-25   06   39.30 &    1.01
$\pm$     0.42 & 149 $\pm$   13
&   13.33$\pm$    0.05&   12.84$\pm$    0.02&   12.75$\pm$    0.01
&  yes \\
  9&   07  14    38.14 &-25   09   42.55 &    1.48
$\pm$     0.21 & 152 $\pm$  04
&   16.44$\pm$    0.02&   15.19$\pm$    0.02&   15.10$\pm$    0.02
&   no \\
 10&   07  14    35.66 &-25   07   49.68 &    0.80
$\pm$     0.31 & 160 $\pm$  08
&   16.58$\pm$    0.03&   15.71$\pm$    0.02&   15.38$\pm$    0.05
&   yes \\
 11&   07  14    33.38 &-25  11    0.84 &    0.63
$\pm$     0.16 & 153 $\pm$  08
&   15.19$\pm$    0.04&   14.57$\pm$    0.01&   14.48$\pm$    0.05
&   yes\\
 12&   07  14    31.80 &-25   09   37.80 &    1.16
$\pm$     0.75 & 156 $\pm$   20
&   16.44$\pm$    0.02&   15.32$\pm$    0.02&   15.03$\pm$    0.02
&   yes\\
 13&   07  14    30.61 &-25   07   20.54 &    1.26
$\pm$     0.32 & 136 $\pm$   07
&   15.93$\pm$    0.04&   14.88$\pm$    0.04&   14.52$\pm$    0.04
&   yes \\
 14&   07  14    30.22 &-25  11   16.24 &    0.91
$\pm$     0.28 & 147 $\pm$   09
&   15.07$\pm$    0.07&   14.61$\pm$    0.02&   14.46$\pm$    0.02
&   yes \\
 15&   07  14    29.71 &-25   08   48.61 &    1.22
$\pm$     0.82 & 163 $\pm$  19
&   17.20$\pm$    0.02&   16.88$\pm$    0.02&   16.72$\pm$    0.02
&   no \\
 16&   07  14    29.54 &-25   09   25.32 &    0.89
$\pm$     0.16 & 148 $\pm$   05
&   17.69$\pm$    0.02&   16.67$\pm$    0.02&   16.19$\pm$    0.02
&   yes \\
 17&   07  14    27.02 &-25   09   20.83 &    0.90
$\pm$     0.16 & 148 $\pm$   05
&   16.06$\pm$    0.06&   14.62$\pm$    0.07&   14.03$\pm$    0.01
&  yes \\
 18&   07  14    26.91 &-25   08   21.61 &    3.06
$\pm$     0.60 & 136 $\pm$   06
&   NA &    NA &   NA
&   --- \\

\hline
\end{tabular}
\label{tab4}
\end{center}
\end{table*}

\begin{table*}
\begin{center}
  \caption{Estimated values of linear polarization in V-band ($\textit{p}_{v}$), polarization angle ($\theta $) and BVR magnitude of various field stars in CB60. }
  \begin{tabular}{|c|c|c|c|c|c||c|c|c|c|}

  \hline

Serial   & RA(J2000)  & Dec.(J2000) &  $\textit{p}_{v}$ & ($\theta $)& B& V  &  R & Main sequence star \\
number   & (h m s)    & $ (^\circ $ $^{\prime}$  $ ^{\prime\prime} $) & $(\%)$ &  $(^\circ)$ &(mag)&(mag)&(mag)&(yes/no)\\
\hline

   1&   08   04    44.76 &-31  31   35.75 &    2.24
$\pm$     0.60 & 166 $\pm$   08
&   17.30$\pm$    0.02&   15.72$\pm$    0.01&   15.20$\pm$    0.02
&   no\\
  2&   08   04    41.56 &-31  32   13.86 &    0.99
$\pm$     0.99 & 155 $\pm$  29
&   16.93$\pm$    0.02&   16.55$\pm$    0.02&   16.06$\pm$    0.02
&   no \\
  3&   08   04    40.20 &-31  29   42.66 &    0.85
$\pm$     0.10 & 147 $\pm$   03
&   13.77$\pm$    0.04&   13.22$\pm$    0.02&   13.07$\pm$    0.02
&   yes\\
  4&   08   04    39.75 &-31  32   51.52 &    0.78
$\pm$     0.28 & 159 $\pm$  10
&   16.47$\pm$    0.01&   15.45$\pm$    0.13&   15.04$\pm$    0.05
& yes \\
  5&   08   04    38.49 &-31  31   32.64 &    0.79
$\pm$     0.05 & 145 $\pm$   02
&   14.17$\pm$    0.04&   13.53$\pm$    0.03&   13.28$\pm$    0.03
&  yes\\
  6&   08   04    38.06 &-31  30   55.98 &    1.23
$\pm$     0.13 & 151 $\pm$   03
&   15.35$\pm$    0.01&   14.59$\pm$    0.10&   14.36$\pm$    0.07
&   yes\\
  7&   08   04    38.07 &-31  32   48.41 &    1.42
$\pm$     0.24 & 152 $\pm$   05
&   16.00$\pm$    0.02&   15.07$\pm$    0.02&   14.56$\pm$    0.02
&   yes \\
  8&   08   04    36.40 &-31  28   54.32 &    0.91
$\pm$     0.21 & 171 $\pm$   07
&   13.13$\pm$    0.03&   12.53$\pm$    0.02&   12.36$\pm$    0.03
&   yes \\
  9&   08   04    36.43 &-31  30   34.96 &    4.10
$\pm$     0.05 & 145 $\pm$   01
&   13.85$\pm$    0.03&   13.34$\pm$    0.03&   13.14$\pm$    0.03
&   no  \\
 10&   08   04    36.17 &-31  31   57.20 &    0.90
$\pm$     0.53 & 165 $\pm$  17
&   17.53$\pm$    0.16&   16.46$\pm$    0.01&   15.63$\pm$    0.02
&   no \\
 11&   08   04    34.91 &-31  31   29.47 &    1.10
$\pm$     0.32 & 147 $\pm$   08
&   17.13$\pm$    0.02&   16.12$\pm$    0.01&   15.64$\pm$    0.01
&   yes \\
 12&   08   04    32.60 &-31  30    01.21 &    1.00
$\pm$     0.39 & 173 $\pm$  11
&   17.28$\pm$    0.02&   16.12$\pm$    0.01&   15.60$\pm$    0.02
&   yes\\
 13&   08   04    30.67 &-31  31   19.33 &    0.60
$\pm$     0.85 & 141 $\pm$  41
&   17.35$\pm$    0.02&   16.56$\pm$    0.02&   15.92$\pm$    0.02
&   no \\

\hline
\end{tabular}
\label{tab5}
\end{center}
\end{table*}

\begin{table*}
\begin{center}
  \caption{Estimated values of linear polarization in V-band ($\textit{p}_{v}$), polarization angle ($\theta $) and BVR magnitude of various field stars in CB69 (NA: Not Available).}
  \begin{tabular}{|c|c|c|c|c|c||c|c|c|c|}
  \hline
Serial   & RA(J2000)  & Dec.(J2000) &  $\textit{p}_{v}$ & ($\theta $)& B& V  &  R & Main sequence star \\
number   & (h m s)    & $ (^\circ $ $^{\prime}$  $ ^{\prime\prime} $) & $(\%)$ &  $(^\circ)$&(mag)&(mag)&(mag)&(yes/no) \\
\hline
  1&  17   02    53.45 &-33  15   57.83 &    0.39
$\pm$     0.08 & 171 $\pm$   06
&   14.58$\pm$    0.02&   12.46$\pm$    0.01&   11.64$\pm$    0.05
&   yes\\
  2&  17   02    52.08 &-33  17   17.30 &    1.37
$\pm$     0.41 & 148 $\pm$   09
&   15.88$\pm$    0.02&   15.42$\pm$    0.02&   15.55$\pm$    0.02
&   yes \\
  3&  17   02    51.51 &-33  18    00.28 &    2.51
$\pm$     0.59 & 137 $\pm$   07
&   16.10$\pm$    0.02&   15.67$\pm$    0.02&   16.14$\pm$    0.02
&  yes\\
  4&  17   02    51.01 &-33  16   41.54 &    1.16
$\pm$     0.31 & 150 $\pm$   08
&   15.64$\pm$    0.02&   15.30$\pm$    0.02&   15.82$\pm$    0.02
&   yes \\
  5&  17   02    47.48 &-33  17   52.97 &    2.31
$\pm$     0.65 & 164 $\pm$   08
&   17.08$\pm$    0.02&   16.45$\pm$    0.02&   17.51$\pm$    0.02
&   no \\
  6&  17   02    46.58 &-33  14   57.93 &    1.23
$\pm$     0.38 & 169 $\pm$   09
&   15.52$\pm$    0.08&   14.50$\pm$    0.04&   14.20$\pm$    0.08
&   yes \\
  7&  17   02    45.96 &-33  18   56.40 &    3.28
$\pm$     0.68 & 146 $\pm$   06
&   NA &    NA &   NA
&   --- \\
  8&  17   02    43.39 &-33  18    00.21 &    0.85
$\pm$     0.06 & 159 $\pm$   02
&   14.58$\pm$    0.06&   13.68$\pm$    0.06&   13.35$\pm$    0.02
&   yes\\
  9&  17   02    40.56 &-33  18   30.62 &    2.73
$\pm$     0.65 & 145 $\pm$   07
&   18.35$\pm$    0.02&   16.60$\pm$    0.02&   16.47$\pm$    0.02
&   yes \\
 10&  17   02    39.99 &-33  15    02.75 &    1.59
$\pm$     0.37 & 140 $\pm$   07
&   15.73$\pm$    0.07&   14.69$\pm$    0.02&   14.30$\pm$    0.04
&   yes\\
 11&  17   02    38.90 &-33  17   37.86 &    2.82
$\pm$     0.52 & 172 $\pm$   05
&   17.92$\pm$    0.02&   16.12$\pm$    0.02&   15.43$\pm$    0.02
&   yes\\
 12&  17   02    38.30 &-33  16   28.08 &    1.40
$\pm$     0.40 & 136 $\pm$   08
&   18.30$\pm$    0.02&   16.04$\pm$    0.02&   14.89$\pm$    0.02
&   yes \\
 13&  17   02    37.05 &-33  16   48.79 &    1.99
$\pm$     0.44 & 172 $\pm$   06
&   16.47$\pm$    0.02&   15.96$\pm$    0.02&   15.86$\pm$    0.02
&   yes \\
 14&  17   02    36.61 &-33  18    08.24 &    2.42
$\pm$     0.43 & 163 $\pm$   05
&   17.78$\pm$    0.02&   15.74$\pm$    0.02&   15.84$\pm$    0.02
&   no\\
 15&  17   02    36.04 &-33  16   56.47 &    4.91
$\pm$     0.67 & 151 $\pm$   04
&   17.08$\pm$    0.02&   16.24$\pm$    0.02&   17.06$\pm$    0.02
&   no \\
 16&  17   02    35.12 &-33  17   35.04 &    1.05
$\pm$     0.77 & 169 $\pm$  21
&   17.98$\pm$    0.02&   16.28$\pm$    0.02&   15.04$\pm$    0.02
&   no \\

\hline
\end{tabular}
\label{tab6}
\end{center}
\end{table*}

\begin{figure}
\begin{center}
\vspace{5 cm}
\includegraphics[width=20pc, height=15pc]{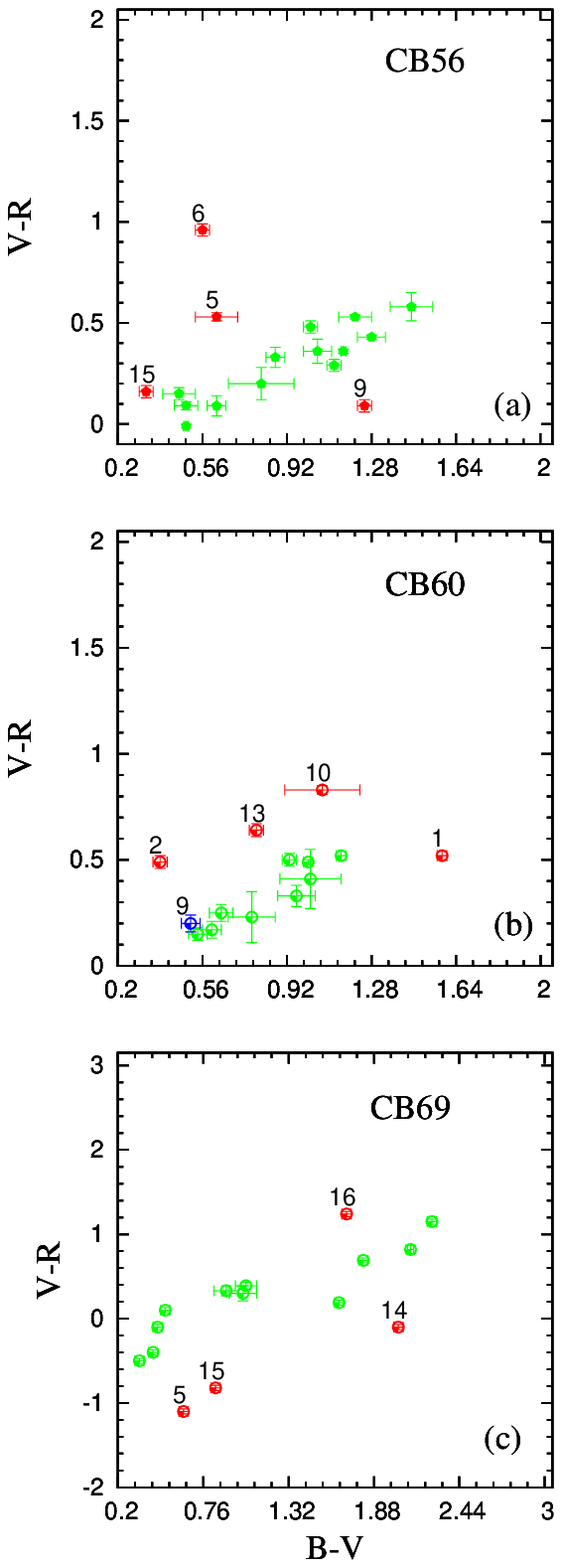}
\caption{ Plot of B-V  versus V-R colour-colour diagram of field stars in the CB56, CB60 and CB69 clouds. Diagonally clustered  points represent stars that are in the main sequence. The stars that do not belong to the main sequence are represented by their serial numbers as mentioned in Tables 4, 5 and 6 (Star \#9 in Figure-4b  is a variable star,  also  known as V517 Pup ).}\label{Fig4}
\end{center}
\end{figure}

Figures \ref{Fig1}, \ref{Fig2} and \ref{Fig3} represent the geometry of the magnetic field, projected on the plane of sky, in the observed field of these clouds, where the polarization vectors and position angles of the respective field stars are superimposed on DSS (Digital Sky Survey) image of each cloud. The length of the  polarization vector is proportional to the degree of polarization, \textit{p}$ _{v} $ ($\%$), and is oriented in the direction given by the position angle $\theta$ (in degree),  measured w.r.t north axis. A reference vector line is drawn with 1$\%$ and 90$^{\circ}$ position angle. The dashed line is superimposed on the polarization map of each cloud, indicating the orientation of the projection of the Galactic plane (GP).

\subsection{Estimation of colour excess}

\cite{b2} described a systematic process to estimate the $E(B-V)$ for a main sequence star, from its observed B, V and R magnitude. Dereddened B, V and R magnitudes are then computed from this colour excess, using a, R$ _{v} $(=A($ \lambda $)/A($ V $)) dependent empirical relationship given by \cite{b5}. This procedure is then repeated up to the convergence to zero of the residual colour excess, which reaches only after a certain number of iterations for each of the field stars. The total colour excess $E(B-V)$ value is then computed. However, the validity and the accuracy of this technique of estimating $E(B-V)$ are discussed in detail in both \cite{b2} and \cite{b141}. To verify the technique, they compared the derived colour excess from this technique with the spectroscopic method and found to be working properly. So in the present work we have adopted the same procedure to estimate the colour excess of main sequence stars of CB56, CB60 and CB69.

In Figure-\ref{Fig4}, (B - V) versus (V - R) color-color diagram along with error bars is plotted for all the field stars of each cloud. It can be inferred from the graph that mostly all the field stars are clustered together along a diagonal locus. The diagonally clustered stars in the colour-colour diagram represent the main-sequence stars, while the scattered ones represent non main sequence stars. They are represented by the serial number as mentioned earlier in Table \ref{tab4}, \ref{tab5} and \ref{tab6}.

R$_{v}(= A _{v}/E(B-V)) $ is an important parameter to characterize dust properties at optical wavelength because it is closely related to the size and composition of dust grains. It is well established that R$_{v}$ has a uniform value of $ \sim $ 3.1 in the diffuse interstellar medium \citep{bwghs}. But it tends to be much higher in dense molecular cloud (4$ \leq $ R$_{v} \leq$6 )  (\cite{bvct}, \cite{bscacp},  \cite{b2}). Since the magnetic field of background starlight of three clouds CB56, CB60 and CB69 has been traced through optical polarimetry which works only in the regions of low extinction,  so our analysis is mainly concentrated on the low-density edge regions of the clouds. For this reason,  we have taken R$_{v}$ = 3.1 and estimated the value of $E(B-V)$ for main sequence stars.


\begin{table*}
\begin{center}
 \begin{minipage}{140mm}
  \caption{ Estimated values of $E(B-V)$, A$ _{v} $  and $ \textit{p}_{v} /A _{v} $  of various field stars in CB56. $E(B-V)$ for Star No. 18 couldn't be calculated due to unavailability of its BVR magnitude.}
  \hspace{-1.5cm}
  \begin{tabular}{|c|c|c|c|c|c|c|c|c|}
  \hline
\footnotetext{A$ _{v} $=R$ _{v} $E(B-V).}
\footnotetext{b background.}
\footnotetext{f foreground.}
\footnotetext{? Not sure}
\footnotetext{unknown: The star does not belong to main sequence, so we couldn't estimate the value of E(B-V).}
Serial   & $E(B-V)$  &  $ A_{v} $   &  $ \textit{p}_{v} /A _{v} $& location \\
number   & ~  & ~  &  ~ & ~&  \\
\hline
1&	0.78$\pm$0.06&	2.42$\pm$	0.19&	0.52$\pm$	0.20& b\\
2&	0.57$\pm$0.07&	1.77$\pm$	0.22&	0.42$\pm$	0.16&?\\
3&	0.92$\pm$0.01&	2.85$\pm$	0.04&	0.36$\pm$	0.30&b\\
4&	0.60$\pm$0.01& 	1.86$\pm$	0.04&	0.68$\pm$	0.18&?\\
5& --- &   --- &   --- &  unknown &  			\\
6& --- &   --- &   --- &  unknown &   				\\
7&	0.63$\pm$0.14&	1.95$\pm$	0.43&	0.41$\pm$	0.14&	b\\
8&	0.47$\pm$0.05 &	1.46$\pm$	0.17&	0.69$\pm$	0.33&	?\\
9& ---&   --- &   --- &    unknown  &   				\\
10&	0.50$\pm$0.04&	1.55$\pm$	0.11&	0.52$\pm$	0.15&	? \\
11&	0.60$\pm$0.04& 	1.86$\pm$	0.13&	0.34$\pm$	0.09&	?\\
12&	0.82$\pm$0.03&	2.54$\pm$	0.09&	0.45$\pm$	0.32&	b\\
13&	0.64$\pm$0.06&	1.98$\pm$	0.18&	0.63$\pm$	0.17&	b\\
14&	0.34$\pm$0.07&	1.05$\pm$   0.22&   0.86$\pm$ 	0.32&   f\\
15&---&  --- &---&    unknown  &     		\\
16&	0.45$\pm$0.03&	1.40$\pm$0.09&  0.63$ \pm $0.12 & f 	\\
17&	0.73$\pm$0.09&	2.26$\pm$	0.29&	0.40$\pm$	0.08&	b\\
18&   --- &   --- &   --- &   unknown&    \\

\hline
\end{tabular}
\label{tab7}
\end{minipage}
\end{center}
\end{table*}

\begin{table*}
\begin{center}
 \begin{minipage}{140mm}
 \caption{ Estimated values of $E(B-V)$, A$ _{v} $  and $ \textit{p}_{v} /A _{v} $  of various field stars in CB60.}
    \hspace{-1.5cm}
   \begin{tabular}{|c|c|c|c|c|c|}
\hline
\footnotetext{A$ _{v} $=R$ _{v} $E(B-V).}
\footnotetext{b background.}
\footnotetext{f foreground.}
\footnotetext{? Not sure}
\footnotetext{unknown: The star does not belong to main sequence, so we couldn't estimate the value of E(B-V).}
Serial   & $E(B-V)$  &$ A_{v} $ & $ \textit{p}_{v} /A _{v} $  & location &\\
number   & ~  & ~  &  ~ & ~&  \\
\hline

1&	--- &   --- &   --- &   unknown  & \\
2&	--- &   --- &   --- &   unknown   \\
3&	0.43$\pm$0.05&	1.33$\pm$0.14	&0.64$\pm$0.08&	?\\
4&	0.55$\pm$0.13&	1.71$\pm$0.40	&0.46$\pm$0.19&	b\\
5&	0.38$\pm$0.05&	1.18$\pm$0.16	&0.67$\pm$0.10&	?\\
6&	0.54$\pm$0.10&	1.67$\pm$0.31	&0.74$\pm$0.16&	b\\
7&	0.33$\pm$0.02&  1.02$\pm$0.06 &1.40$\pm$0.25& f\\
8&	0.46$\pm$0.04&	1.43$\pm$0.11	&0.64$\pm$0.16&	?\\
9&	--- &	--- &   --- &  variable star &   \\
10&	---  &   --- &   --- &  unknown &  				\\
11&	0.43$\pm$0.02&	1.33$\pm$0.07	&0.82$\pm$0.24&	?\\
12&	0.53$\pm$0.02&	1.64$\pm$0.07	&0.61$\pm$0.24&	b\\
13&	--- &   --- &   --- &   unknown &  \\							

\hline
\end{tabular}
\label{tab8}
\end{minipage}
\end{center}
\end{table*}

\begin{table*}
\begin{center}
\begin{minipage}{140mm}
 \caption{Estimated values of $E(B-V)$, A$ _{v} $  and $ \textit{p}_{v} /A _{v} $  of various field stars in CB69. $E(B-V)$ for Star No. 7 couldn't be calculated due to unavailability of its BVR magnitude.}
   \hspace{-1.5cm}
  \begin{tabular}{|c|c|c|c|c|c|c|c|c|}
\hline
\footnotetext{A$ _{v} $=R$ _{v} $E(B-V).}
\footnotetext{b background.}
\footnotetext{f foreground.}
\footnotetext{? Not sure}
\footnotetext{unknown: The star does not belong to main sequence, so we couldn't estimate the value of E(B-V).}
Serial   & $E(B-V)$  &$ A_{v} $ & $ \textit{p}_{v} /A _{v} $  & location &\\
number   & ~  & ~  &  ~ & ~&  \\
\hline

1&	1.09$\pm$0.02&	3.38$\pm$0.07&	0.12$\pm$0.03&	b\\
2&	0.77$\pm$0.03&	2.39$\pm$0.09&	0.57$\pm$0.17&	?\\
3&	1.40$\pm$0.03&	4.34$\pm$0.09&	0.58$\pm$0.14&	b\\
4&	1.40$\pm$0.03&	4.34$\pm$0.09&	0.27$\pm$0.07&	b\\
5&		---& --- &	 ---	   &  unknown				\\
6&	0.70$\pm$0.09& 	2.17$\pm$0.28&	0.57$\pm$0.19&	?\\
7&		---& --- &	---	    & unknown			\\
8&	0.54$\pm$0.09& 	1.67$\pm$0.26&	0.51$\pm$0.09&	?\\
9&	1.83$\pm$0.03&	5.67$\pm$0.09&	0.48$\pm$0.12&	b\\
10&	0.59$\pm$0.07& 	1.83$\pm$0.23&	0.86$\pm$0.23&	?\\
11&	0.96$\pm$0.03&	2.98$\pm$0.09&	0.95$\pm$0.18&	b\\
12&	0.81$\pm$0.03& 	2.51$\pm$0.09&	0.56$\pm$0.16&	?\\
13&	0.46$\pm$0.03&	1.43$\pm$0.09&	1.39$\pm$0.32&	f\\
14&		& ---	&	    ---     &	unknown		\\
15&		& ---	&	    ---	    &	unknown		\\
16&		& ---	&	    ---	    &	unknown		\\

\hline
\end{tabular}
\label{tab9}
\end{minipage}
\end{center}
\end{table*}

Table 7, 8 and 9 show the calculated values of $E(B-V)$, $A_{v}$, polarization efficiency ($ \textit{p}_{v} $/A$ _{v} $) and probable location of the main sequence field stars of CB56, CB60 and CB69 respectively.

\subsubsection{Location identification}	
To identify the location of different field stars in a cloud we use statistical approach and define: $l = [E(B-V)]_{avg} - \sigma$, where $[E(B-V)]_{avg}$ is the average extinction of the cloud and $\sigma$ is the standard deviation of E(B-V). If E(B-V) of any star is less than $l$, then it may be considered as foreground to the cloud. The star whose $E(B-V)$ is more than $l + \sigma$ is considered background whereas  the value within $l$ and $l + \sigma$ is considered either within the cloud or maybe background to the cloud.

For CB56, the value of $r^{2}$ (coefficient of determination computed between $E(B-V)$ and $p_v$ along the observed line of sight) is 0.14, which gives the proportion of the variance (fluctuation) of polarization that can be predicted from the $E(B-V)$. It means the polarization of the starlight along the observed line of sight is weakly correlated with the extinction. The average extinction and standard deviation are $[E(B-V)]_{avg}$ = 0.62 (mag) and $\sigma$ = 0.16. In Table 7, we have categorized the observed field stars of CB56, in three distinct groups on the basis of their location. Stars with serial no. 14 and 16, have $E(B-V) < l$ which may be located foreground to the cloud. Thus excluding these two stars the average value of $E(B-V)$ of rest 11 stars is 0.66 (mag) and the value of $r ^{2}  = 0.13$. On the other hand, stars with serial no. 1, 3, 7, 12, 13 and 17 have $E(B-V) > l + \sigma$, so they may be considered to be located background of the cloud. However, the $E(B-V)$ of stars with serial no. 2, 4, 8, 10,and 11, lies within $l$ and $l + \sigma$, so they may be either within the cloud or located background to the cloud. Thus, these stars fall into third group whose location cannot be claimed with certainty.  Star with serial no. 18 couldn't be located as the BVR magnitude of this star is not available in  the VizieR database. Stars with serial no. 5, 6, 9 and 15 (please see table \ref{tab4}) are not in main sequence class so we have excluded them from present analysis.

In the observed field of CB60, star having serial no. 9 (in Table 5) is identified to be variable star (V517 Pup) which is confirmed from SIMBAD database. So we have excluded this from our present analysis. The average extinction and standard deviation are given by $[E(B-V)]_{avg}$ = 0.46 (mag) and $\sigma$ = 0.08. The $r ^{2}$ is estimated to be 0.09. Table 8 indicates the location of observed field stars of CB60 which are categorized following the argument as discussed earlier for identifying the location of the field stars of CB56. Excluding the foreground star (serial no. 7), the mean $E(B-V)$ is 0.48 (mag), and the value of $r ^{2} $ is 0.11. Stars with serial no. 1, 2, 10 and 13 (please see table \ref{tab5}) are not in main sequence class so we have excluded them from present analysis.
	
Table 9 displays the location for the observed field stars of CB69. The average extinction and standard deviation of CB69 field stars are $[E(B-V)]_{avg}$ = 0.96 (mag) and $\sigma$ = 0.43 where $r ^{2}  =0.15$ . We  identified the location of some field stars of this cloud, one foreground star (13), five background stars (1, 3, 4, 9 and 11) and five stars within the cloud (2, 6, 8, 10, and 12). The average $E(B-V)$ excluding the foreground star  is 1.01 (mag), and the value of $r ^{2} $ is 0.24. However, the location of star number 7 couldn't be identified as the BVR magnitude is not available in the VizieR database of astronomical catalogues. Stars with serial no. 5, 14, 15 and 16 (please see table \ref{tab5}) are not in main sequence class so we have excluded them from present analysis.

\subsection{Morphology of the magnetic field}
	\citet{b23} have probed 248 small molecular clouds (mostly Bok globules), using deep IRAS photometry and $^{12}$CO spectroscopy. Comparing their CO peak line temperature and CO line width they have divided those clouds into three types A, B, and C. The A type cloud contains the maximum number of clouds (74$\%$ of 248 clouds) where gas temperature are cool ( $\sim$8.5K) and with little turbulent gas motion. The B type cloud has unusually warm temperature, but with narrow line width and the C type cloud has  quite a broad line width, implying unusual dynamic activity, whereas the gas temperature are cool.  It is observed from \citet{b23} that cloud CB56 and CB69 are identified as A type cloud, and CB60 as B type cloud.

Figures 5a--5f represent the distribution of \textit{p} and $\theta$ in the V-band, for the field stars observed in the Bok globules: CB56, CB60 and CB69.

\begin{figure*}
\begin{center}
\vspace{7 cm}
\includegraphics[width=25pc, height=17pc]{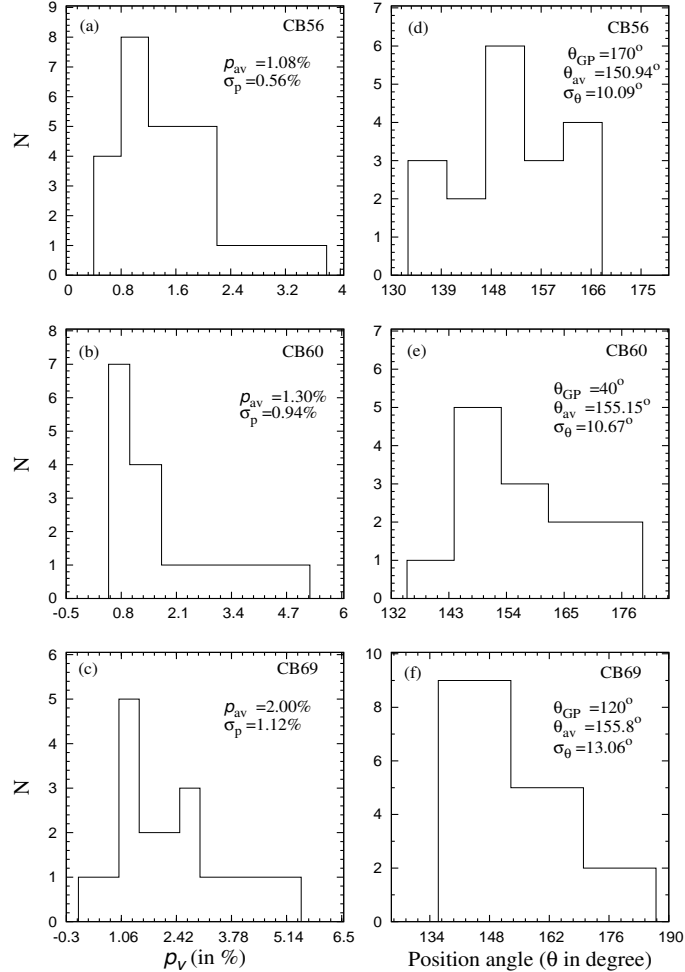}
\caption{Histogram shows the distribution of degree of linear polarization and  polarization position angle vs N (Number of star) for CB56, CB60, and CB69 Bok globules.}
\label{Fig5}
\end{center}
\end{figure*}

 Cloud CB56 has two IRAS sources (07125-2503; 07125-2507) where its major axis has a position angle of 170$^{\circ}$ (\citet{b4}). The mean value for polarization and their position angle for the CB56 field stars are $ p_{av} $ = 1.08$ \% $, and $\theta_{av}$= 150.94$ ^{\circ} $ with a standard deviation of $ \sigma_{p}$ = 0.56$ \% $ and $ \sigma_{\theta} $ = 10.09$^{\circ} $ respectively. The polarization vector of most of the stars is aligned in a common direction which indicates the local magnetic field of the cloud. The alignment of polarization vectors in local magnetic field is quite expected as A type clouds show less dynamical activities and turbulence. Thus, they have better alignment of the grains. It is further noticed from Fig. 1 that the direction of the local magnetic field of the cloud is almost parallel to the galactic plane. Thus we can infer that the field direction in this cloud is similar to that of the galactic field. Further, significant variation in the polarization value of star 18 (please see Table 4) from mean polarization value, suggests that it may be located either within the cloud or maybe located background to the cloud. It is worth mentioning that BVR magnitudes are not available for this star, so it is not possible to comment its location from photometry.

 From figure \ref{Fig2}, it can be seen that the polarization vectors in CB60 (B-Type cloud) seem to be directed in some common direction of the local magnetic field of the cloud. The direction of local field ($\theta_{av} = 155.15^{\circ}$) is not aligned in the direction of GP (40$^{\circ}$) which represents that the magnetic field within the cloud is different from the galactic field.  Also it is to be noted that B type cloud has some unusual warm temperature. It has three IRAS sources (08026-3122; 08029-3118; 08022-3155) (\citet{b4}). The average polarization and their position angle of the CB60 field stars are $ p_{av} $ =1.30$ \% $, and $\theta_{av}$= 155.15$ ^{\circ} $ with a standard deviation of $ \sigma_{p}$ = 0.94$ \% $ and $ \sigma_{\theta} $ = 10.67$^{\circ}$ respectively. The polarization value of star 9 is unusually high from mean polarization value (please see Table 5), which is also a known variable star V517 Pup.

It can be observed from figure \ref{Fig3} that for the cloud CB69 (A type cloud) the polarization vectors are well aligned but the general direction is not along the direction of the GP ($\theta _G = 120^{\circ}$). This cloud has six IRAS sources (16594-3315, 16595-3311, 16586-3304, 16589-3315, 16590-3313, 16590-3305) (\citet{b4}). The average polarization and their position angle of the CB60 field stars are $ p_{av} $ = 2.00$ \% $, and $\theta_{av}$= 155.8$ ^{\circ} $ with a standard deviation of $ \sigma_{p}$ = 1.12$ \% $ and $ \sigma_{\theta} $ = 13.06$^{\circ}$ respectively.

The degree of linear polarization varies (lower/higher) depending on the location of the stars in the cloud. Similar variation (alignment/non-alignment) of the  polarization position angle with GP are generally observed depending on the membership of the stars (i.e. foreground/background/within the cloud). Thus, the efficiency of polarization depends on both, the properties of dust grain and on the environment in which they exist. So the study of variation of the background star polarization with extinction (A$ _{v} $) and polarization efficiency (defined as $\textit{p}_{v} $/A$ _{v} $), can help us in understanding the nature of the dust and the magnetic field associated with the dark cloud (\citet{b10}, \cite{b1}).

\subsection{Polarization efficiency}
The alignment of non spherical dust grains are generally measured by the ratio of degree of polarization produced for a given amount of extinction ($ \textit{p}_{v} $/A$ _{v} $), also known as the \emph{polarization efficiency} \citep{b150}.  \citet{bsmf} observed the linear polarization of 180 nearby stars in  U, B, V and R bands where they studied the relation of the maximum polarization $ \textit{p}_{max}$ with the color excess $E(B-V)$ at a wavelength $ \lambda_{max}$, and found that the polarization efficiency along different line of sight varies within an upper limit: $p_{max}$/E(B-V)$\leq$ 9.0 $\%$ mag$ ^{-1}$.

\begin{figure}
\begin{center}
\vspace{2 cm}

\includegraphics[width=30pc, height=20pc]{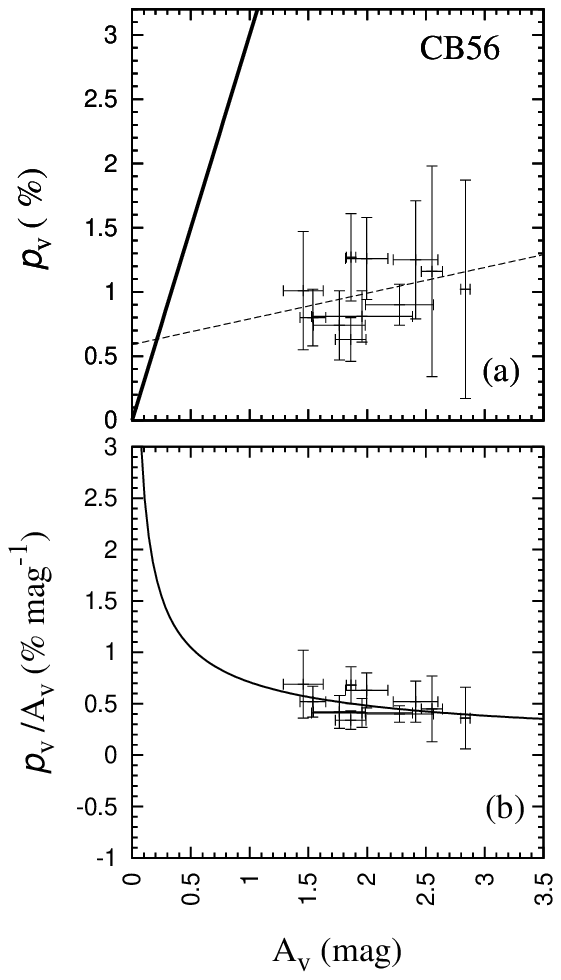}
\\
\caption{(a) Upper panel is the plot of A$ _{v} $ vs $ \textit{p}_{v} $ (where $R_v$ = 3.1). The solid line in the figure represents upper limit of polarization efficiency ($ \textit{p}_{v} $=3.0 A$ _{v} $), and the dotted line is the  unweighted best fit line of the field stars of CB56 .
(b) Lower panel is the plot of A$ _{v} $ vs $ \textit{p}_{v} $/A$ _{v} $ (where $R_v$ = 3.1). The solid line is the unweighted best fit power law ($ p_{v}/A_{v} = 0.71 A_{v}^{-0.56} $) for the background field stars of CB56. }\label{Fig6}
\end{center}
\end{figure}

\begin{figure}
\begin{center}
\vspace{2 cm}
\includegraphics[width=30pc, height=20pc]{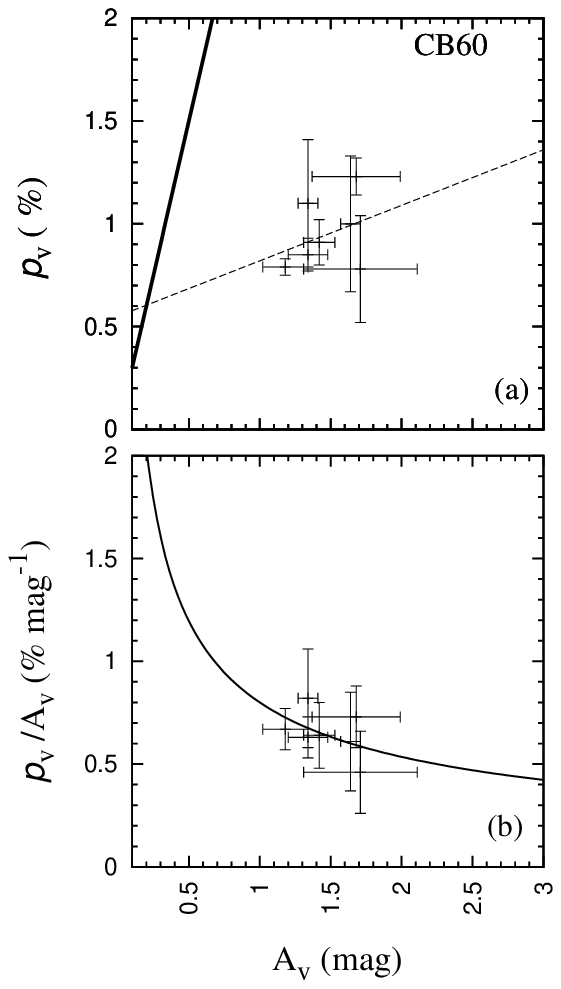}

\caption{(a) Upper panel is the plot of A$ _{v} $ vs $ \textit{p}_{v} $ (where $R_v$ = 3.1). The solid line in the figure represents upper limit of polarization efficiency ($ \textit{p}_{v} $=3.0 A$ _{v} $), and the dotted line is the unweighted best fit line of the field stars of CB60 .
(b) Lower panel is the plot of A$ _{v} $ vs $ \textit{p}_{v} $/A$ _{v} $ (where $R_v$ = 3.1). The solid line is the  unweighted best fit power law ($ \textit{p}_{v}/A_{v}=0.81 A_{v}^{-0.59} $) for the background field stars of CB60.}\label{Fig7}
\end{center}
\end{figure}

\begin{figure}
\begin{center}
\vspace{1 cm}
\includegraphics[width=30pc, height=20pc]{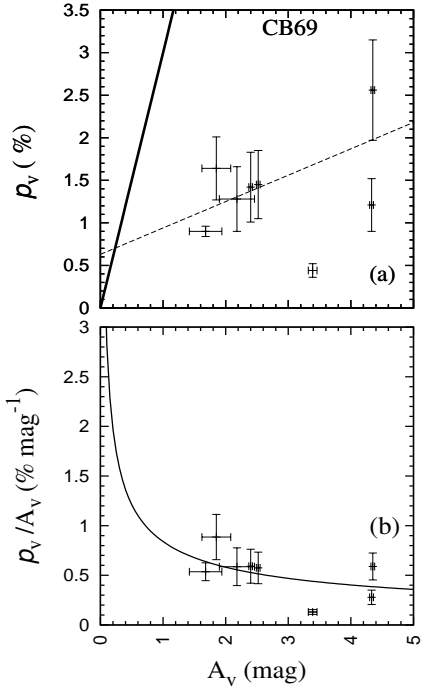}

\caption{(a) Upper panel is the plot of A$ _{v} $ vs $ p_{v} $ (where $R_v$ = 3.1). The solid line in the figure represents upper limit of polarization efficiency ($p_{v} $=3.0 A$ _{v} $), and the dotted line is the unweighted best fit line of the field stars of CB69 .
(b) Lower panel is the plot of A$ _{v} $ vs $ p_{v}$ $/A_{v} $ (where $R_v$ = 3.1). The dotted line is the unweighted best fit power law ($ p_{v}/A_{v} = 0.84 A_{v}^{-0.52}$) for the background field stars of CB69.}\label{Fig8}
\end{center}
\end{figure}

\citet{b150}  calculated the theoretical upper limit of the polarization efficiency of the grains due to selective extinction which is $\leq$ 14$\%$ mag$^{-1}$. However, the observational upper limit of polarization efficiency, is found to be $ \sim $ 3$ \% $mag$ ^{-1} $ a factor of four less than the predicted value for the ideal scenario \citep{b150}. Figure \ref{Fig6} (a), \ref{Fig7}(a) and \ref{Fig8}(a) are the plot of $A_v $ vs $ \textit{p}_{v} $ for the three clouds CB56, CB60 and CB69. It can be seen that the value of $A _{v}$ for all the stars lies below the line, drawn by using the relation: $\textit{p}_{v} (\%)= 3 A _{v}$ (mag).


  Several investigators (\citet{btnst}, \citet{b8}, \citet{b9}, \citet{b10} and \citet{bwth}) in past showed that the efficiency of polarization  decrease systematically with the increase in extinction and it can be represented by a power law $\textit{p}_{v}$/A$_{v}\varpropto A ^{-\alpha}_{v} $ for some index $\alpha$. Figure \ref{Fig6} (b), \ref{Fig7}(b) and \ref{Fig8}(b) are the plots of A$_v$ vs $ \textit{p}_{v}/ $ A$_v $ for the three clouds CB56, CB60 and CB69. The unweighted power law fits to all the background field stars in the cloud CB56, CB60 and CB69 are given by

\begin{equation}
 p_{v}/A_{v}=(0.71 \pm 0.18)A_{v}^{-0.56 \pm 0.36} ~{\%}~  mag^{-1}
 \end{equation}

\begin{equation}
 p_{v}/A_{v}=(0.81 \pm 0.16)A_{v}^{-0.59 \pm 0.51} ~{\%}~  mag^{-1}
 \end{equation}

\begin{equation}
 p_{v}/A_{v}=(0.84 \pm 0.48)A_{v}^{-0.52 \pm 0.49} ~{\%}~  mag^{-1}
 \end{equation}

These power law explain the dependence of polarization efficiency on visual extinction. Our calculated power laws are consistent with the findings of \citet{b10} and \citet{bwth} for TDC. It is to be noted that our analysis is based on V-band data set  whereas their findings are based on  K-band data set. It may be concluded that the linear polarization of the starlight due to aligned dust grains in these three clouds is more efficient in regions of low extinction as compared to high obscured lines of sight. Thus the characteristics of material composing these Bok globules are consistent with the material found in diffuse interstellar medium (ISM). However present analysis is restricted to limited number of data points.

\section{CONCLUSIONS}

\begin{enumerate}
\item Three Bok globules (CB56, CB60 and CB69) have been probed at the V-band, in order to map the magnetic field within the observed region of the cloud. The distribution of $p$ and $\theta$ in the V-band are studied for these clouds. It has been found that the polarization vectors of most of the stars in these globules are aligned in some common direction which indicate the direction of local magnetic field in the globules. The local magnetic field of the cloud CB56 ($\textit{l}$ = 237.92 $ ^{\circ}$, $\textit{b} = - 6.46 ^{\circ}$) is almost aligned with the galactic field whereas the alignment is not observed in CB60 ($\textit{l}$ = 248.89 $ ^{\circ}$, $\textit{b} = - 0.01^{\circ}$) and CB69 ($\textit{l}$ = 351.23 $ ^{\circ}$, $\textit{b}$ = 5.14$^{\circ}$).

\item  $E(B-V)$ for the observed field stars are calculated from the observed BVR magnitudes. These values are estimated for  $R _{v}  $ = 3.1. From the estimated $E(B-V)$, the probable location of each field stars is identified.

\item  The polarization efficiency of these clouds is also studied, which shows decrease in polarization efficiency with an increase in extinction along the observed line of sight. This indicates that the non spherical dust grains in these globules, is consistent with dust grains found in the diffuse interstellar medium. It is to be noted that the decrease in polarization could, in some cases, result from a change in the direction of the magnetic field along the line of sight, not necessarily a decrease in alignment efficiency. Also power laws calculated for CB56, CB60 and CB69 are in good agreement with the findings of both \citet{b10} and \citet{bwth} for TDC.

\end{enumerate}

\section*{ACKNOWLEDGMENTS}

We gratefully acknowledge IUCAA, Pune for making telescope time available.  The reviewers of this paper are highly acknowledged for their constructive comments which definitely helped us to improve the quality of the paper.  We also acknowledge the use of the VizieR database of astronomical catalogues namely UCAC4  \citep{b18}, NOMAD \citep{b17}, DENIS database \citep{b19}. This work is supported by the Department of Science \& Technology (DST), Government of India, under SERC-Fast Track scheme (Dy. No. SERB/F/1750/2012-13).

\label{lastpage}


\begin{thebibliography}{99}

\bibitem[\protect\citeauthoryear{Alves et al. }{2008}]{bafg}
Alves, F. O., Franco, G. A. P., \& Girart, J. M., 2008, A\&A, 486, L13

\bibitem[\protect\citeauthoryear{Arce et al. }{1998}]{b1}
Arce, H. G., Goodman, A. A., Bastein, P., Manset, N., \& Summer, M., 1998, ApJ, 499, L93

\bibitem[\protect\citeauthoryear{Bernabei \& Polcaro}{2001}]{b2}
Bernabei, S., \& Polcaro, V. F., 2001, A\&A, 371, 123

\bibitem[\protect\citeauthoryear{Bok \& Reilly et al. }{1947}]{b3}
Bok, B. J., \& Reilly, E. F., 1947, ApJ, 105, 255

\bibitem[\protect\citeauthoryear{Bhatt \& Jain et al. }{1993}]{bbj}
Bhatt, H. C. \& Jain, S. K., 1993, A\&A, 276, 507

\bibitem[\protect\citeauthoryear{Clemens \& Barvainis et al. }{1988}]{b4}
Clemens, D. P., \& Barvainis R., 1988, ApJS, 68, 257

\bibitem[\protect\citeauthoryear{Clemens et al. }{1991}]{b23}
Clemens, D. P., Yun J. L., Heyer M. H., 1991, ApJS, 75, 877

\bibitem[\protect\citeauthoryear{Cardelli et al. }{1989}]{b5}
Cardelli, J. A., Clayton, G. C., \& Mathis, J. S., 1989, ApJ, 345, 245

\bibitem[\protect\citeauthoryear{Das et al. }{2013}]{bdmwbdc}
Das, H. S., Medhi, B. J., Wolf, S., Bertrang, G., Deb Roy, P., Chakraborty, A., 2013, MNRAS, 436, 3500

\bibitem[\protect\citeauthoryear{Consortium }{2005}]{b19}
The DENIS Consortium, 2005, VizieR Online Data Catalog, 2263, 0

\bibitem[\protect\citeauthoryear{Franco et al. }{2010}]{bfag}
Franco, G. A. P., Alves, F. O., \& Girart, J. M., 2010, ApJ, 723, 146

\bibitem[\protect\citeauthoryear{Goodman et al. }{1990}]{bgj}
Goodman, A. A., Bastien, P., Menard, F., \& Myers, P. C., 1990, ApJ, 359, 363

\bibitem[\protect\citeauthoryear{Goodman et al. }{1992}]{b8}
Goodman, A. A., Jones, T. J., Lada, E. A., \& Myers, P. C., 1992, ApJ, 399, 108

\bibitem[\protect\citeauthoryear{Goodman et al. }{1995}]{b9}
Goodman, A. A., Jones, T. J., Lada, E. A., \& Myers, P. C., 1995, ApJ, 448, 748

\bibitem[\protect\citeauthoryear{Gerakines et al. }{1995}]{b10}
Gerakines, P. A., Whittet, D. C. B., \& Lazarian, A., 1995, ApJ, 455, L171

\bibitem[\protect\citeauthoryear{Jones}{1989}]{jtj}
Jones, T. J., 1989, ApJ, 346, 728

\bibitem[\protect\citeauthoryear{Jones et al. }{1992}]{jkd}
Jones, T. J., Klebe, D., Dickey, J. M. 1992, ApJ, 389, 602


\bibitem[\protect\citeauthoryear{Kane et al. }{1995}]{b11}
Kane, B. D., Clemens, D. P., Leach, R. W., \& Barvainis, R., 1995, ApJ, 445, 269


\bibitem[\protect\citeauthoryear{Launhardt et al. }{1997}]{blh}
Launhardt, R., \&  Henning, Th., 1997, A\&A, 326, 329

\bibitem[\protect\citeauthoryear{Lazarian et al. }{1997}]{blgm}
Lazarian A., Goodman Alyssa A., and  Myers Philip C., 1997, ApJ, 490, 273

\bibitem[\protect\citeauthoryear{Myers et al. }{1991}]{b22}
Myers, P. C., Goodman A. A., 1991, ApJ, 373, 509

\bibitem[\protect\citeauthoryear{Paul et al. }{2012}]{b12}
Paul, D., Das, H. S., \& Sen, A. K., 2012, BASI, 40, 113

\bibitem[\protect\citeauthoryear{Ramaprakash et al.}{1998}]{b16}
Ramaprakash, A.N., Gupta Ranjan, Sen A.K., Tandon S.N., 1998, A\&AS 128, 369

\bibitem[\protect\citeauthoryear{Sen et al. }{2000}]{b13}
Sen, A. K., Gupta, R., Ramprakash, A. N., \& Tandon, S. N. 2000, A\&AS, 141, 175

\bibitem[\protect\citeauthoryear{Sen et al. }{2010}]{b141}
Sen, A. K., Polcaro, V.F., Dey., I, \& Gupta, R., 2010, A\&A, 522, A45

\bibitem[\protect\citeauthoryear{Serkowski et al. }{1975}]{bsmf}
Serkowski, K., Mathewson, D. L., \& Ford, V. L., 1975, ApJ, 196, 261

\bibitem[\protect\citeauthoryear{Strafella et al. }{2001}]{bscacp}
Strafella, F., Campeggio, L., Aiello, S., Cecchi-Pestellini, C. \& Pezzuto, S., 2001, ApJ, 558,
717

\bibitem[\protect\citeauthoryear{Tamura et al. }{1987}]{btnst}
Tamura, M., Nagata, T., Sato, S., \& Tanaka, M., 1987, MNRAS, 224, 413.

\bibitem[\protect\citeauthoryear{Vrba et al. }{1976}]{bvss}
Vrba, F. J., Strom, S. E., \& Strom, K. M., 1976, AJ, 81, 958

\bibitem[\protect\citeauthoryear{Vrba et al. }{1977}]{bv}
Vrba, F. J., 1977, AJ, 82, 198

\bibitem[\protect\citeauthoryear{Vrba, Coyne & Tapia }{1981}]{bvct}
Vrba, F. J., Coyne, G. V., \& Tapia, S., 1981, ApJ, 243, 489

\bibitem[\protect\citeauthoryear{Vrba et al. }{1986}]{bvls}
Vrba, F. J., Luginbuhl, C. B., Strom, S. E., Strom, K. M., \& Heyer, M. H., 1986,
AJ, 92, 633

\bibitem[\protect\citeauthoryear{Vrba et al. }{1993}]{bvct}
Vrba, F. J., Coyne, G. V., \& Tapia, S., 1993, AJ, 105, 1010

\bibitem[\protect\citeauthoryear{Ward-Thompson et al.}{2009}]{bwskn}
Ward-Thompson, D., Sen, A. K., Kirk, J. M., \& Nutter, D., 2009, MNRAS, 398, 394

\bibitem[\protect\citeauthoryear{Whittet et al.}{2001}]{bwghs}
Whittet, D. C. B., Gerakines, P . A., Hough, J. H., \& Shenoy , S. S., 2001, ApJ, 547, 872

\bibitem[\protect\citeauthoryear{Whittet et al.}{2003}]{b150}
Whittet, D. C. B., ed., 2003, Dust in the galactic environment

\bibitem[\protect\citeauthoryear{Whittet et al.}{2008}]{bwth}
Whittet, D. C. B., Hough, J. h., Lazarian, A., \& Hoang , T., 2008, ApJ, 674, 304


\bibitem[\protect\citeauthoryear{Yun et al.}{1994}]{byc}
Yun, J. L., \& Clemens, D, P., 1994, AJ, 108, 2

\bibitem[\protect\citeauthoryear{Zacharias et al.}{2005}]{b17}
Zacharias N., Monet D. G., Levine S. E., Urban S. E., Gaume R., Wycoff G. L., 2005, VizieR Online Data Catalog, 1297, 0.

\bibitem[\protect\citeauthoryear{Zacharias et al.}{2012}]{b18}
Zacharias, N., Finch C.T., Girard T.M., Henden A., Bartlet J.L., Monet D.G., Zacharias M.I., 2012, 2013, AJ, 145, 44.







\end{thebibliography}
\end{document}